\documentstyle[pra,aps,amsfonts]{revtex}
\begin{document}
\draft
\twocolumn[\hsize\textwidth\columnwidth\hsize\csname 
@twocolumnfalse\endcsname
\title{Conservation of Orbital Angular Momentum in Stimulated
Down-Conversion} 
\author{D. P. Caetano, M. P. Almeida  and P.H. Souto Ribeiro$^{*}$}
\address{Instituto de F\'{\i}sica, Universidade Federal do Rio de 
Janeiro, Caixa Postal 68528, Rio de Janeiro, RJ 22945-970, Brazil}
\author{J. A. O. Huguenin, B. Coutinho dos Santos and A. Z. Khoury}
\address{Instituto de F\'{\i}sica,  Universidade Federal Fluminense, 
BR-24210-340 Niteroi, RJ, Brazil} 
\date{\today}
\maketitle
\begin{abstract}
We report on an experiment demonstrating the conservation of orbital angular momentum
in stimulated down-conversion. The orbital angular momentum is not transferred to the
individual beams of the spontaneous down-conversion, but it is conserved when twin photons 
are taken individually. We observe the conservation law for an individual beam of the
down-conversion through cavity-free stimulated emission.
\end{abstract}
\pacs{42.50.Ar, 42.25.Kb}
]
The cavity-free stimulated parametric down conversion was first studied by Mandel
and co-workers\cite{1,2} and more recently it has been explored by other authors\cite{3,4,5}.
One important aspect of this process is its connection with the spontaneous parametric
down-conversion, where entangled states for two photons can be easily prepared.
Signals obtained in stimulated down-conversion are much larger than those obtained in the
spontaneous process and carries information about the details of the parametric interaction,
like phase matching conditions. This information is preserved thanks to the stimulation
without optical cavities, where the optical mode properties are determined mainly by the
cavity configuration. 
Therefore, studying stimulated down-conversion is useful for understanding
entanglement properties of the twin photons from the parametric down-conversion. We have
recently demonstrated the transfer of coherence and images from the pump and auxiliary lasers
to the stimulated down-conversion field\cite{5} in direct connection with the analogous process in
the context of the quantum correlations observed in coincidence measurements\cite{7}.

The possibility of preparing entangled photons in different degrees of freedom, has also
become subject of interest. Particularly, the orbital angular momentum (OAM) of the light,
has been studied in the context of the classical\cite{9} and quantum optics\cite{10}.
Conservation of OAM in the up-conversion process\cite{11}, optical pumping of
cold atoms\cite{12} and quantum entanglement\cite{10} have been observed
experimentally  for this degree of freedom. 
However, in the spontaneous parametric down-conversion process, the OAM
is not transferred from the pump to each individual signal or idler beam\cite{13}. This is a
consequence of the fact that signal and idler beams are incoherent when taken individually\cite{14}.

In this work, we observe experimentally the manifestation of the conservation law for the
OAM in the stimulated down-conversion process, for the idler beam. In this case, besides 
the pump, a second auxiliary laser is aligned with one of the down-conversion modes inducing
emission. Conservation  of the topological charge, can be written as
m$_{p}$ = m$_{s}$ + m$_{i}$, where p,s,i stands for pump, signal and idler respectively.

Light beams with OAM can be described by Laguerre-Gauss LG$_{l,m}$ modes, where $l$ and
$m$ are azimuthal and radial mode numbers, and the OAM is given by $m\hbar$ per photon.
In our experiment, LG$_{0,1}$ modes are produced by diffraction on computer generated holograms
as in Ref.\cite{15}, for example.
The identification of the modes, was made in our experiment
by the passage of each beam through a Michelson interferometer, operating with
a small misalignment\cite{15}. The resulting interference pattern shows the sign and the absolute 
value of  the topological charge in the mode.
This method is very simple and presents some advantages compared to other most common ones,
where a coherent reference field is needed, or where a Dove prism is inserted inside a Mach-Zhender
interferometer. 

The spatial intensity distribution of the idler beam in the stimulated down-conversion for
thin crystals can be predicted by Eq. 10 of Ref. \cite{4}. Special cases, where spontaneous emission
is negligible and the transverse amplitude of one of the laser fields can be considered constant, 
result in Eqs. 2 and 3 of Ref. \cite{5}. From these equations it is seen that the intensity
profile of the idler beam will look like a doughnut if the pump or the auxiliary laser is
prepared in a LG$_{0,m}$($m\neq 0$) mode.This is indeed an indication that the idler beam is also a 
LG$_{0,m}$ mode,
but rigorously this is not enough. In Refs. \cite{4,5}, the quantum treatment used has shown to be
useful in describing the stimulated down-conversion process and it would be interesting
to derive the state of the idler field when either the pump or the auxiliary laser is prepared in LG
modes, within the same formalism. However, this calculation is not straightforward and it is beyond the 
scope of the present work. In the following, we will present experimental results supporting the predictions in Ref.\cite{4} concerning the intensity distributions and supporting the intuition that if
the idler presents a doughnut shape it "should" posses some OAM.

\begin{figure}[h]
\vspace*{5.5cm}
\special{eps: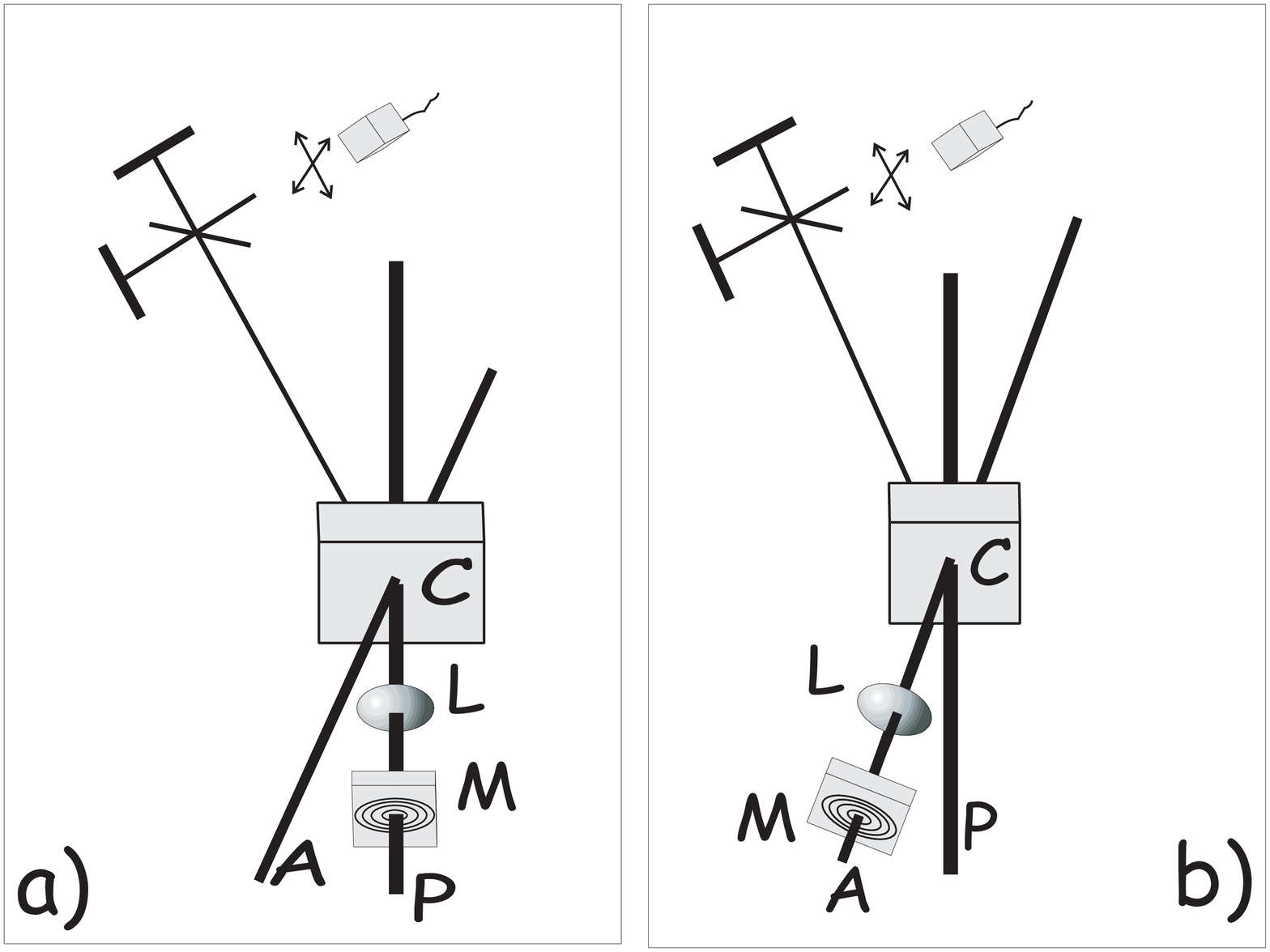 x=8.5cm y=6cm}
\caption{Sketch of the experiment. P is the pump beam, A is the auxiliary beam, L is the lens, C is the nonlinear crystal, M is a diffraction mask. a) The pump beam is prepared in the LG$_{0,1}$ mode.
b) The auxiliary beam is prepared in the LG$_{0,1}$ mode.}
\label{fig1}
\end{figure}

The experimental set-up is sketched in Fig. \ref{fig1}. A He-Cd laser pumps a BBO non-linear
crystal 3mm long, with a c.w. 442 nm wavelength beam. Non-degenerate twin beams with
signal and idler wavelengths around 845nm and 925nm respectively, are generated. An auxiliary
beam is obtained from a diode laser oscillating around 845nm. It is aligned with the signal beam,
so that their modes have good overlap and emission is stimulated in this down-conversion mode
by the laser. As a result, the idler beam is completely changed with respect to its intensity
and spectral properties, as described in Refs.\cite{1,2,3,4,5}. The goal of the experiment is
to prepare the pump beam in a LG$_{0,1}$ mode and to measure the OAM of the
idler beam. The same procedure is repeated preparing the auxiliary beam in a LG$_{0,1}$ mode
and measuring the OAM of the idler beam. The idler beam is directed onto a Michelson interferometer, before it is detected by an avalanche photodiode single photon counting module. The Michelson interferometer is slightly misaligned along the horizontal axis, so that for a plane wave input, the resulting interference pattern presents vertical parallel stripes. The larger the misalignment,
the narrower the stripes. When a LG$_{0,m}$ ($m \neq 0$) mode enters
the interferometer, the beam with doughnut shape is divided in two and the misalignment works to 
make the side of one beam interfere with the center of the other and vice-versa. Two opposed bifurcations appear in the interference pattern. The orientation of the bifurcations are
related to the sign of the topological charge, or the sense of rotation of the phase in the
transverse plane and the number of derivations in the fork is related to the absolute value of the
charge.

\begin{figure}[h]
\vspace*{3.5cm}
\special{eps: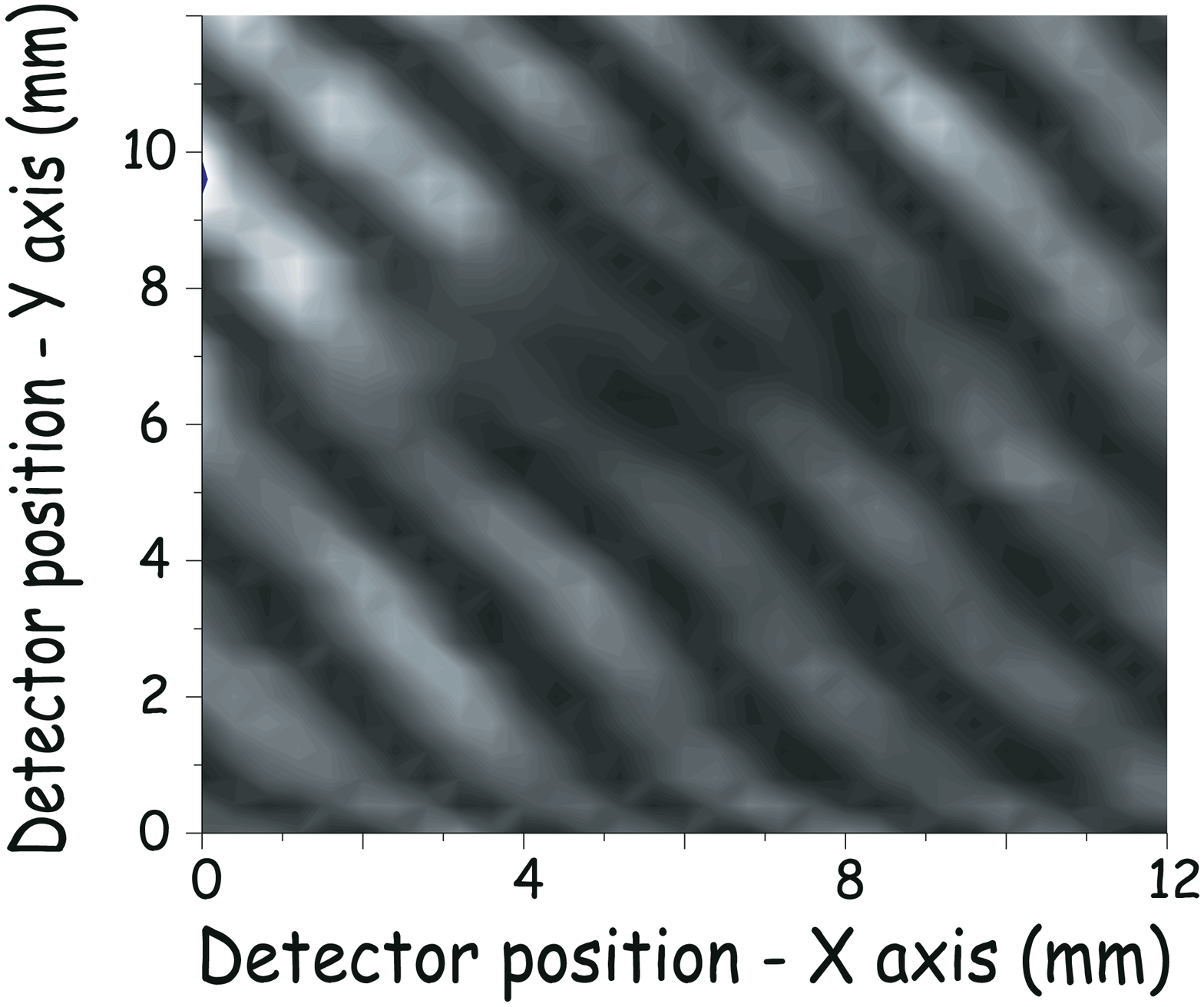 x=5cm y=4cm}
\caption{Gray scale bitmap plotted from a 30x30 matrix with the transverse
interference pattern of the pump beam.}
\label{fig2}
\end{figure}

The pump beam was prepared in a LG mode with m$_{p}$=+1.
After crossing the crystal, the beam is directed to a Michelson interferometer, in the same
fashion as described above for the idler, in order to be able to compare the interference patterns
for pump and idler. 
All interference patterns are measured by scanning  the detector in the transverse plane.
The resulting matrix with the intensities at different positions is converted into a
grey scale bitmap where the higher intensities are white and the lower ones are black.
The interference pattern measured for the pump beam is shown in Fig. \ref{fig2}. 
The two forks would be oriented along the vertical axis if the misalignment were only in the 
horizontal direction.
Due to a small vertical misalignment\cite{16}, the forks are oriented along an axis making an 
angle with the vertical direction. From the orientation of the forks it is possible to identify
the topological charge, m$_{p}$=+1.
As a consequence of the OAM conservation, when the pump is prepared with m$_{p}$=+1 
and the auxiliary  laser with m$_{s}$=0, the idler must have m$_{i}$=+1. On the other hand, 
when the auxiliary laser is prepared with m$_{s}$=+1 and the pump with m$_{p}$=0, 
the idler must have m$_{i}$= -1.

\begin{figure}[h]
\vspace*{3.5cm}
\special{eps: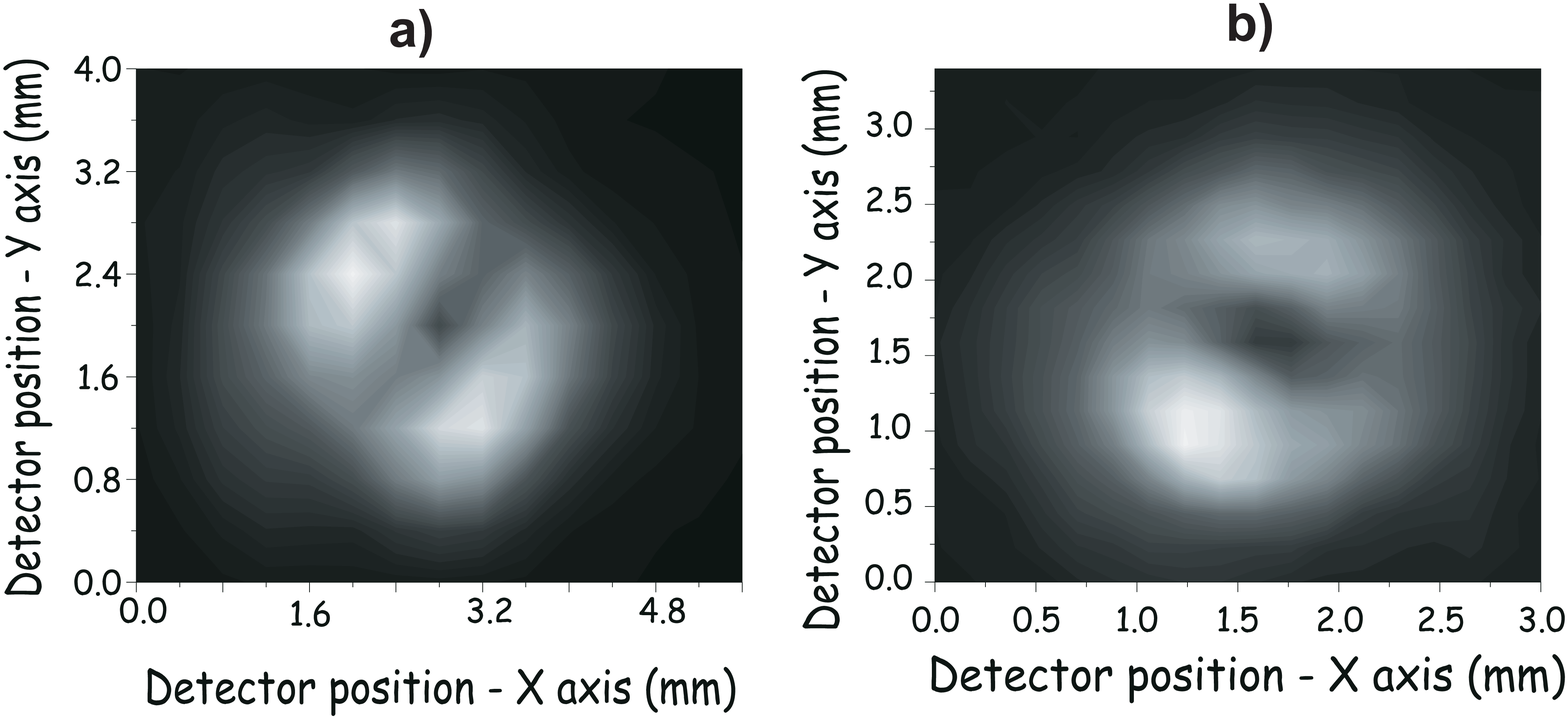 x=8.5cm y=4cm}
\caption{Gray scale bitmap plotted from a 20x20 matrix with the transverse
intensity of the idler beam. a)Pump LG$_{0,1}$ mode m$_{p}$=+1 and b)Auxiliary LG$_{0,1}$ mode m$_{s}$=+1}
\label{fig3}
\end{figure}

\begin{figure}[h]
\vspace*{5.5cm}
\special{eps: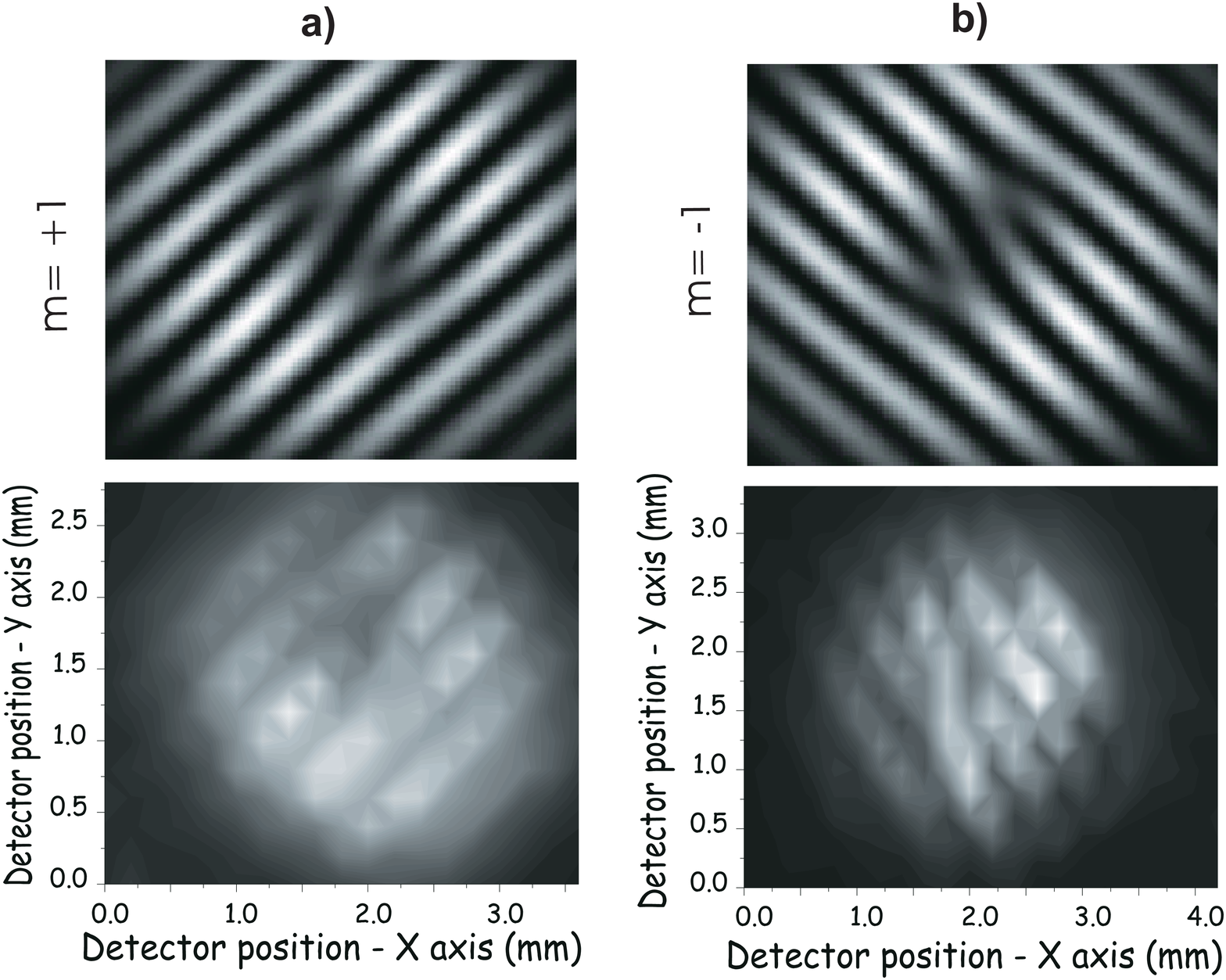 x=8.5cm y=6cm}
\caption{Gray scale bitmap plotted from a 20x20 matrix with the transverse
interference pattern of the idler beam. a)Pump LG$_{0,1}$ mode m$_{p}$=+1. Theoretical
simulation (top), experimental result(bottom). b)Auxiliary LG$_{0,1}$ mode m$_{s}$=+1.
Theoretical simulation (top), experimental result(bottom).}
\label{fig4}
\end{figure}

The idler beam obtained in the stimulated down-conversion is then analyzed with the
Michelson interferometer in the same way as described above for the pump. Pump and auxiliary
laser powers are high enough to ensure that the spontaneous emission is negligible
compared to the stimulated one. As a result, when either the pump or the auxiliary laser
is prepared in a LG$_{0,m}$($m\neq 0$) mode, the idler beam also propagates as a 
LG$_{0,m}$($m\neq 0$) mode and its intensity distribution looks like a doughnut. 
Idler intensity distributions are shown in Fig.\ref{fig3}, when
a) m$_{p}$=+1 and m$_{s}$=0 and b) m$_{p}$=0 and m$_{s}$=+1. 
In Fig.\ref{fig4} we have the interference patterns, again  for a) m$_{p}$=+1 and m$_{s}$=0, 
and b) m$_{p}$=0 and m$_{s}$=+1, where the upper plots correspond to theoretical
simulations and the lower ones correspond to the experimental results for the idler beam.
Note that the orientation of the forks of the idler in a) (m$_{i}$=+1) is inverted (mirror image)
when compared to the idler in b) (m$_{i}$= -1).

In the results presented above, the OAM was actually transferred from the pump
and auxiliary lasers to the stimulated idler beam. When the OAM comes from the
pump with m$_{p}$=+1, the idler is changed into a m$_{i}$=+1 LG$_{0,1}$ mode. When
the OAM comes from the auxiliary lasers with m$_{s}$=+1, the idler is changed
into a m$_{i}$= -1 LG$_{0,-1}$ mode. This is compatible with the conservation of the
total topological charge m$_{p}$ = m$_{s}$ + m$_{i}$. Since pump, signal and idler
beams are not collinear, the conservation of the OAM requires that part of the
momentum is absorbed by the crystal.
The  relation, m$_{i}$ = - m$_{s}$ when m$_{p}$ = 0,
can be understood in terms of the phase conjugation of the idler in comparison with
the auxiliary laser\cite{5}, as a LG beam with m=+1 looks like a LG beam with m= -1
propagating backwards.

In conclusion, we have observed experimentally the transfer of orbital angular
momentum from the pump and auxiliary lasers to the stimulated parametric
down-conversion idler beam. This transfer implies in the conservation of the
topological charge. For concluding that the orbital angular momentum vector is conserved 
it is necessary the crystal absorbing part of the momentum, since
pump, signal and idler beams are not collinear.

The authors thanks Dr. D. Petrov and Dr. P.A.M. dos Santos for helping in
manufacturing holographic masks and Dr. C. H. Monken for fruitful discussions.
Financial support was provided by Brazilian agencies CNPq, PRONEX, CAPES, FAPERJ 
and FUJB.

\begin{thebibliography}{99}
\bibitem{16}
The interference pattern for the pump beam can be easily controlled,
so that we have intentionally introduced a small vertical misalignment
in order to demonstrate its effect. For the idler beam it is harder to
control as it is a weak infra-red beam.
\end{thebibliography}
\end{document}